\documentclass[10pt,conference]{IEEEtran}
\IEEEoverridecommandlockouts

\usepackage{cite}

\usepackage{amsmath,amssymb,amsfonts}
\usepackage{algorithmic}
\usepackage{graphicx}
\usepackage{textcomp}
\usepackage{url} 
\usepackage{xcolor}
\usepackage{colortbl}
\usepackage{pgfplots}
\usepackage{tikz}
\usepackage[framemethod=TikZ]{mdframed}
\definecolor{lightyellow}{RGB}{255,255,224}
\usepackage{tcolorbox}
\usepackage{multirow}
\usepackage{color, soul}

\def\BibTeX{{\rm B\kern-.05em{\sc i\kern-.025em b}\kern-.08em
    T\kern-.1667em\lower.7ex\hbox{E}\kern-.125emX}}
 \makeatletter
\newcommand{\linebreakand}{%
  \end{@IEEEauthorhalign}
  \hfill\mbox{}\par
  \mbox{}\hfill\begin{@IEEEauthorhalign}
}
\makeatother  

\makeatletter
\def\ps@IEEEtitlepagestyle{
  \def\@oddfoot{\mycopyrightnotice}
  \def\@evenfoot{}
}
\def\mycopyrightnotice{
  {\footnotesize
  \begin{minipage}{\textwidth}
  \begin{mdframed}
       \copyright~2025 IEEE. Personal use of this material is permitted. Permission from IEEE must be obtained for all other uses, in any current or future media, including reprinting/republishing this material for advertising or promotional purposes, creating new collective works, for resale or redistribution to servers or lists, or reuse of any copyrighted component of this work in other works. To be published in the Proceedings of the 6th ACM/IEEE Workshop on Gender Equality, Diversity, and Inclusion in Software Engineering (GE), 2025 in Ottawa, Ontario, Canada.
  \end{mdframed}

  \end{minipage}
  }
}
\begin{document}

\title{Overcoming Obstacles: Challenges of Gender Inequality in Undergraduate ICT Programs\\
}

\author{\IEEEauthorblockN{
Angelica Pereira Souza}
 \IEEEauthorblockA{
\textit{State University of Bahia}\\
 Alagoinhas, Brazil \\
  angelicapereirasouza@hotmail.com}
  \and
  \IEEEauthorblockN{
  Anderson Uchôa}
 \IEEEauthorblockA{
  \textit{Federal University of Ceará}\\
  Itapajé, Brazil \\
  andersonuchoa@ufc.br}
  \and
  \IEEEauthorblockN{
  Edna Dias Canedo}
  \IEEEauthorblockA{
  \textit{University of Brasília (UnB)}\\
  Brasilia, Brazil \\
  ednacanedo@unb.br}
 \linebreakand
  \centering
  \IEEEauthorblockN{
  Juliana Alves Pereira}
  \IEEEauthorblockA{
  \textit{Pont. Catholic Univ. of Rio de Janeiro}\\
  Rio de Janeiro, Brazil \\
  juliana@inf.puc-rio.br}
  \and
\IEEEauthorblockN{
Claudia Pinto Pereira
  \IEEEauthorblockA{
  \textit{State Univ. of Feira de Santana}\\
  Feira de Santana, Brazil \\
  claudiap@uefs.br}
  \and
\IEEEauthorblockN{
Larissa Rocha}
  \IEEEauthorblockA{
  \textit{State University of Bahia}\\
  Alagoinhas, Brazil \\
 larissabastos@uneb.br}}
 
  }

\maketitle

\begin{abstract}
Context: Gender inequality is a widely discussed issue across various sectors, including Information Technology and Communication (ICT). In Brazil, women represent less than 18\% of ICT students in higher education. Prior studies highlight gender-related barriers that discourage women from staying in ICT. However, they provide limited insights into their perceptions as undergraduate students and the factors influencing their participation and confidence.
Goal: This study explores the perceptions of women undergraduate students in ICT regarding gender inequality.
Method: A survey of 402 women from 18 Brazilian states enrolled in ICT courses was conducted using a mixed-method approach, combining quantitative and qualitative analyses.
Results: Women students reported experiencing discriminatory practices from peers and professors, both inside and outside the classroom. Gender stereotypes were found to undermine their self-confidence and self-esteem, occasionally leading to course discontinuation.
Conclusions: Factors such as lack of representation, inappropriate jokes, isolation, mistrust, and difficulty being heard contribute to harmful outcomes, including reduced participation and reluctance to take leadership roles. Addressing these issues is essential to creating a safe and respectful learning environment for all students.
\end{abstract}


\section{Introduction}

The number of students interested in Information and Communication Technology (ICT) courses has increased globally in recent years~\cite{zweben2019taulbee}.
In 2021, there was a significant rise in student enrollment in Brazil, escalating from 295,257 in 2020 to 464,269 (+57.25\%). It can be the result of an increase of 51 05\% in ICT courses since 2010, increasing to 3,015 in 2021 \cite{INEP}. 
Despite the significant increase in overall enrollment, women represented only a small fraction; out of 464,269 students, only 16.3\% were women.
Along these lines, in The USA, only $\sim$20\% of bachelor's degrees in computer science were earned by women in 2018~\cite{nationalScienceFoundation}.
 
Understanding the reasons behind this gender inequality discrepancy requires a more in-depth analysis of gender dynamics within the ICT field. Previous studies explore the factors influencing student's choice of their major degree and investigate why the gender distribution varies significantly across different courses~\cite{doi:10.1080/08993408.2022.2160144, campbell2022stereotypes, cheryan2017some}. Although awareness of the existence of programs may increase the likelihood of their selection, this awareness alone cannot fully account for explaining gender segregation in ICT courses. This phenomenon is influenced by a more intricate set of factors encompassing students' perceptions and experiences within broader sociocultural contexts.
For example, Lamolla and Ramos~\cite{doi:10.1080/13668803.2018.1483321} mentioned several factors that hinder women's entry into the ICT field. These include the lack of inspiring role models, inadequate support in workplaces, and the perception of the ICT environment as male-dominated and overly aggressive in terms of self-confidence.  
 
To illustrate how these factors interact and perpetuate gender disparities, Cheryan et al.~\cite{cheryan2017some} proposed an analogy that illustrates how the barriers perceived and experienced by women can be likened to the reluctance to enter a pool with cold water. The metaphor highlights three points: \textit{(i)} The men's culture as cold water - the predominance of men's culture in the ICT field can be decisive for women, just as the cold temperature of the water can deter someone from entering a pool; \textit{(ii)} The first experience of jumping into the pool - representing the moment of entry or attempt to enter the field. If the initial experience is positive, this can encourage continuation and development. Conversely, a negative experience can reinforce perceived barriers and difficulties and discourage persistence; and \textit{(iii)} Making the ``water temperature" comfortable for everyone - creating more inclusive and welcoming environments in fields dominated by men, 
encouraging participation from a broader context of people, regardless of gender.

The challenges faced by women in the ICT field have been broadly investigated in the literature, mainly in the industrial setting~\cite{10.1145/3613372.3613394, 10164701}.
Most of the studies report that women experience feelings of non-belonging and discrimination, lack of representation, gender stereotypes that diminish women's capabilities, lack of support from society, and reports of lack of support from colleagues.
In complement to prior studies, we aim to investigate perceived gender-related challenges in the academic environment. So, we surveyed more than 400 women and identified the consequences of studying in an environment dominated by men. 
Additionally, we explored the women's motivations, identified possible factors that can influence their career choices, and pointed out practices to improve women’s confidence in tech from their perspective.

The findings reveal that women students feel unwelcome and inferior in classrooms.
Many of them are constantly interrupted by 
men colleagues and professors in academic spaces. These interruptions impact women's self-esteem and lead to discouragement. More than 90\% of respondents affirm that gender stereotypes and lack of representation reduce self-confidence and make them question whether they are capable of understanding ICT-related subjects and exercising a profession in the area. 
Our results suggest a notable correlation between the number of women professors and increased levels of confidence among women students in expressing their views and actively engaging in discussions.
We also highlighted strategies they mentioned to boost their confidence, such as initiatives that can contribute to their enrollment and permanence in ICT programs 
and policies for protection against moral harassment.

\section{Background and Related Work}\label{back&related}


\textbf{Navigating the Path: Exploring Students Choices in Major Selection.} Opps and Yadav~\cite{opps2022belongs} conducted a study with 95 middle school students on their perceptions of computer scientists. The results suggest significant differences between women's and men's perceptions of these professional appearances, which may influence students' early interest in the field. Barrett et al.~\cite{doi:10.1080/08993408.2022.2160144} performed semi-structured interviews with 19 students from computer science and bioinformatics courses to gain deeper insights into the factors influencing their choice of major. Their findings indicate that the lack of early exposure for women affects their perception of belongingness and choice.
Although previous studies~\cite{opps2022belongs, doi:10.1080/08993408.2022.2160144}
shed light on the crucial role of early exposure and work opportunities in influencing women's decision to pursue computer science, our research seeks to understand the experiences, challenges, and factors that contribute to the retention and success of hundreds of women who have already decided to enroll in ICT courses. 

\textbf{Persisting in the Course: Enhancing Women's Retention and Belonging in ICT.}
Retention in computer science majors among underrepresented groups, such as women, remains a significant challenge~\cite{DBLP:conf/fie/HolandaASBOKC21}.
The concept of a woman's sense of belonging within academic environments has received significant attention in educational research~\cite{DBLP:conf/fie/HolandaASBOKC21}.
Campbell‐Montalvo et al.~\cite{campbell2022stereotypes} conducted interviews with 55 students from underrepresented minority groups to explore the challenges they face in undergraduate engineering programs. 
The authors found that stereotyping and differential treatment from peers, especially toward women and black students, can severely undermine their feelings of belonging. 
It can influence their overall engagement, performance, and persistence in their chosen fields of study~\cite{campbell2022stereotypes}.
While Campbell‐Montalvo et al.~\cite{campbell2022stereotypes} provide a broader view of underrepresented groups in engineering, our study delves deeper into the specific challenges of women in ICT programs. We seek to understand not just the barriers to belongingness, but also the strategies women employ to overcome these challenges.

\textbf{Beyond Graduation: Assessing Career Opportunities for Women in ICT.}
The significance of studies and interventions focusing on women in computing courses is underscored by the demand for innovative and inclusive workforces 
within technology companies.
This highlights that gender inequality extends beyond academic concerns~\cite{DBLP:journals/cacm/Vardi18c}
Perdriau et al.~\cite{perdriau2024diversity} conducted semi-structured interviews with 11 third and fourth-year undergraduate computer science women. They explore the impact of the diversity-hire narrative (\textit{i.e.}, women are hired more for diversity quotas than for their merit) on their academic and professional journey. The authors show that it can have profound implications for how women perceive their place and legitimacy 
in the industry. Thus, the implications of this study extend beyond the academic sphere into the hiring practices and workplace cultures within the tech industry.
In contrast, our study focuses 
on the entire spectrum of women academic engagement and challenges.

\section{Study Settings}
\label{settings}


Our study goal
is: \textit{analyse} the perception of women; \textit{for the purpose of} understanding what factors influence their participation and performance; \textit{with respect to} (1) classroom interactions, (2) challenges faced, (3) impostor syndrome, (4) underrepresentation of women, and (5) strategies to enhance women's confidence; \textit{from the viewpoint of} women as undergraduate students; \textit{in the context of} Brazilian undergraduate ICT courses. 
Our research questions (RQs) are:~\footnote{Further details and artifacts can be found on our replication package~\cite{pereira_souza_2024_11053838}.}RQ$_1$. What is the \textbf{impact of classroom interactions} on the perception of competence and confidence of woman ICT students?
 RQ$_2$. What are the \textbf{challenges} that woman undergraduate students face in ICT courses? 
RQ$_3$. Is there anything specific in the general \textbf{context} of a computing degree that you believe could be adjusted to \textbf{improve women’s confidence} in technology?

We posed three RQ$_s$ to explore factors that might influence women students’ levels of participation and confidence during their academic lives. In this context, \textbf{RQ$_1$} aims to assess how classroom interactions impact the competence and confidence of women ICT students, aiming to identify factors that promote or hinder their sense of belonging and academic performance in the field. \textbf{RQ$_2$} aimed at identifying the challenges faced by women ICT students. Elicit these challenges is crucial for understanding and addressing barriers to women's success in ICT fields. 
Finally, \textbf{RQ$_3$} aims to identify specific aspects within the ICT area that could be employed or adjusted to enhance women's confidence in technology. By identifying these aspects, we can actively create a more inclusive and women-friendly ICT environment. 

\paragraph{Study Steps and Procedures}\label{subsec:steps} To answer our $RQs$, we use the survey 
guidelines proposed by Linaker et al.~\cite{linaker2015guidelines} for setting up and conducting our study,
as follows. \textbf{Step 1: Defining objectives for information collection.} As the first step, we conducted a brainstorming session to define the goals and scope of the survey. Our main goal is to understand the factors that influence the participation and performance of women in ICT courses. Our second goal is to provide empirically-driven actionable insights to avoid gender inequality in ICT courses; 
\textbf{Step 2: Identifying the target population and sampling.} Our target population was women who completed or are completing courses in the ICT field in Brazil. To ensure the accuracy of our target audience, we incorporated two control questions in the survey: one to confirm the gender of the participants and another to verify the specificity of the course they are attending or have completed, ensuring that we only receive responses from women who are enrolled or have finished an ICT major. 
We also employed the procedures of snowballing sampling~\cite{kitchenham2002principles}, in which initial participants help to identify and recruit additional participants, leading to the expansion of the participant network in an iterative manner; and \textbf{Step 3: Designing survey instrument.} All authors collectively designed an exploratory survey composed of 38 questions (Q): thirty are closed questions, and eight are open questions.
Table~\ref{tab:survey} overviews our survey.

\begin{table}[ht!]
 \centering
 \tiny
 \caption{Survey Questions}
 \label{tab:survey}
 \begin{tabular}{|p{0.5cm}|p{7.3cm}|}
 \hline \textbf{ID} & \textbf{Question} \\ \hline 
 \multicolumn{2}{|c|}{\cellcolor[HTML]{EFEFEF}\textbf{Profile}} \\ \hline
Q1 & How do you identify in relation to your gender identity?
\\ \hline 
Q2 & How do you identify in relation to your sexual orientation?
 \\ \hline 
Q3 & How do you declare yourself?
\\ \hline 
Q4 & What age group are you in?
\\ \hline 
Q5 & In which state do you currently reside?
\\ \hline
\multicolumn{2}{|c|}{\cellcolor[HTML]{EFEFEF}\textbf{Regarding the undergraduate course}} \\ \hline
Q16 & What motivated you to choose a course in the ICT area?
 \\ \hline
Q6 & What course are you currently studying or have you studied?
\\ \hline 
Q7 & What type of institution do/did you study/studied?
 \\ \hline 
Q8 & What is the year you joined the institution?
\\ \hline 
Q9 & How many students entered your class (total number)?
 \\ \hline 
Q10 & How many women were there in your class, including you?
 \\ \hline 
Q11 & Are you aware of how many of your class dropped out?
 \\ \hline 
Q12 & How many 
women professors do/have you had on your course?
 \\ \hline 
 \multicolumn{2}{|c|}{\cellcolor[HTML]{EFEFEF}\textbf{Perceptions}} \\ \hline
Q13 & At what age were you first exposed to technology in a significant way?
 \\ \hline 
Q14 & About the following statement: boys have access to video games, cell phones or other technology equipment at an earlier age than girls.
 \\ \hline
Q15 & If you agreed with the previous question, do you consider that this difference in initial exposure to technology between genders can influence perception and interest in the area of technology?
 \\ \hline
\multicolumn{2}{|c|}{\cellcolor[HTML]{EFEFEF}\textbf{Experience as a student in an ICT course}} \\ \hline
Q17 & In my experience, I have been the target of comments or suggestions that imply that computing courses or related areas are more appropriate for boys than girls.
 \\ \hline
Q18 & Expressing my opinions during a class is usually?
 \\ \hline
Q19 & If you answered "(Very) Uncomfortable", could you share some specific reasons or situations that contributed to that feeling? 
 \\ \hline
Q20 & About the following statement: I feel more comfortable asking questions or expressing my opinion when the professor is a woman.
 \\ \hline
Q21 & About the following statement: I've stopped giving my opinion during a discussion on a topic, even though I knew I was right.
 \\ \hline
Q22 & In the classroom, have you ever been interrupted while giving your opinion or asking a question?
 \\ \hline
Q23 & If the answer to the previous question is anything other than "Never", who were you interrupted by?
 \\ \hline
Q24 & Could you share any specific situation or context in which this interruption occurred and whether it had an impact on your self-esteem and/or confidence?
 \\ \hline
Q32 & Which of these challenges have you experienced as an undergraduate?
 \\ \hline
Q25 & I've been through a situation in the technology field that made me feel like I was a failure or doubted my abilities and achievements.
 \\ \hline
Q26 & If the answer to the previous question is other than "I disagree", could you say more about that moment?
 \\ \hline
Q27 & Have you ever caught yourself thinking that your achievements were the result of other factors rather than your own skills and efforts?
 \\ \hline
Q28 & If your answer was anything other than "Never", you believe/believed that your achievements were the result of: 
 \\ \hline
Q29 & Do you constantly worry about being "discovered" as a fraud, even when you have solid evidence of your academic success?
 \\ \hline
Q30 & Do you hesitate to accept praise or recognition for your work, feeling that you do not deserve the praise you receive?
 \\ \hline
Q31 & Do you often compare yourself to your 
men colleagues and believe they are more capable or deserving of success than you?
 \\ \hline
\multicolumn{2}{|c|}{\cellcolor[HTML]{EFEFEF}\textbf{Trustness}} \\ \hline
Q33 & Do you believe that gender stereotypes in the area of Information and Communication Technology (ICT) influence this lack of confidence that some women may experience? Comment on it.
 \\ \hline
Q34 & Is there anything specific in the general context of a computing degree that you believe could be adjusted to improve women's confidence in technology?
 \\ \hline
\multicolumn{2}{|c|}{\cellcolor[HTML]{EFEFEF}\textbf{Initiatives}} \\ \hline
Q35 & Is there any initiative (program, networking groups, mentoring, etc.) at the institution where you study or studied regarding awareness about diversity and gender equality in computing? If so, what is the name of the initiative?
 \\ \hline
Q36 & Do you participate or have you participated in any such initiative? \\ \hline
\multicolumn{2}{|c|}{\cellcolor[HTML]{EFEFEF}\textbf{Additional experiences}} \\ \hline
Q37 & Are there any additional comments or suggestions you would like to share about your experiences as a computing major?
 \\ \hline
\end{tabular}
\end{table}

At the beginning of the survey, we provided study information, including the objective, methodology, how the data would be handled, and researchers' contact information. Additionally, we presented the informed consent statement outlining the conditions and stipulations governing participation.\footnote{The conditions adhered to ethical privacy standards as per the Brazilian General Data Protection Law (Law No. 13,709/2018).} In this context, participants were informed that their participation was entirely voluntary, as well as the freedom to not participate or withdraw their consent at any time without suffering penalties. The survey was anonymous and respondents were not asked for any contact information. 
The first section aimed to collect the women's profiles. To this end, we defined a set of five questions. The second section focused on the undergraduate course, including eight questions. The subsequent sections explored different aspects of the students' experiences in ICT courses. The third section addressed personal perceptions with three questions, while the fourth section investigated more detailed student experiences, using sixteen questions. The fifth section examined the participants' confidence through two questions. The sixth section aimed to identify existing initiatives in institutions, focused on diversity and gender equality in ICT, with two questions. Finally, the seventh section was dedicated to collecting additional experiences from the participants, allowing them to share comments or suggestions about their academic journeys.

\textbf{Step 4: Validating the survey.}  We conducted a pilot test with three ICT students who were not involved in the study. Based on the pilots, we improved the wording of some questions. Regarding the time required to complete the questionnaire, the pilot participants took, on average, 15 minutes. We informed respondents of this average time when the questionnaire was made publicly available.
\textbf{Step 5: Sending the survey.} We hosted our survey on Google Forms and shared it through cards and text on social media. We used posts and direct messages on Twitter, LinkedIn, Facebook, Instagram, and WhatsApp. The survey was available for 33 days, from March 1 to April 2, 2024.

\begin{figure}[ht!]
     \centering
\includegraphics[width=.8\columnwidth]{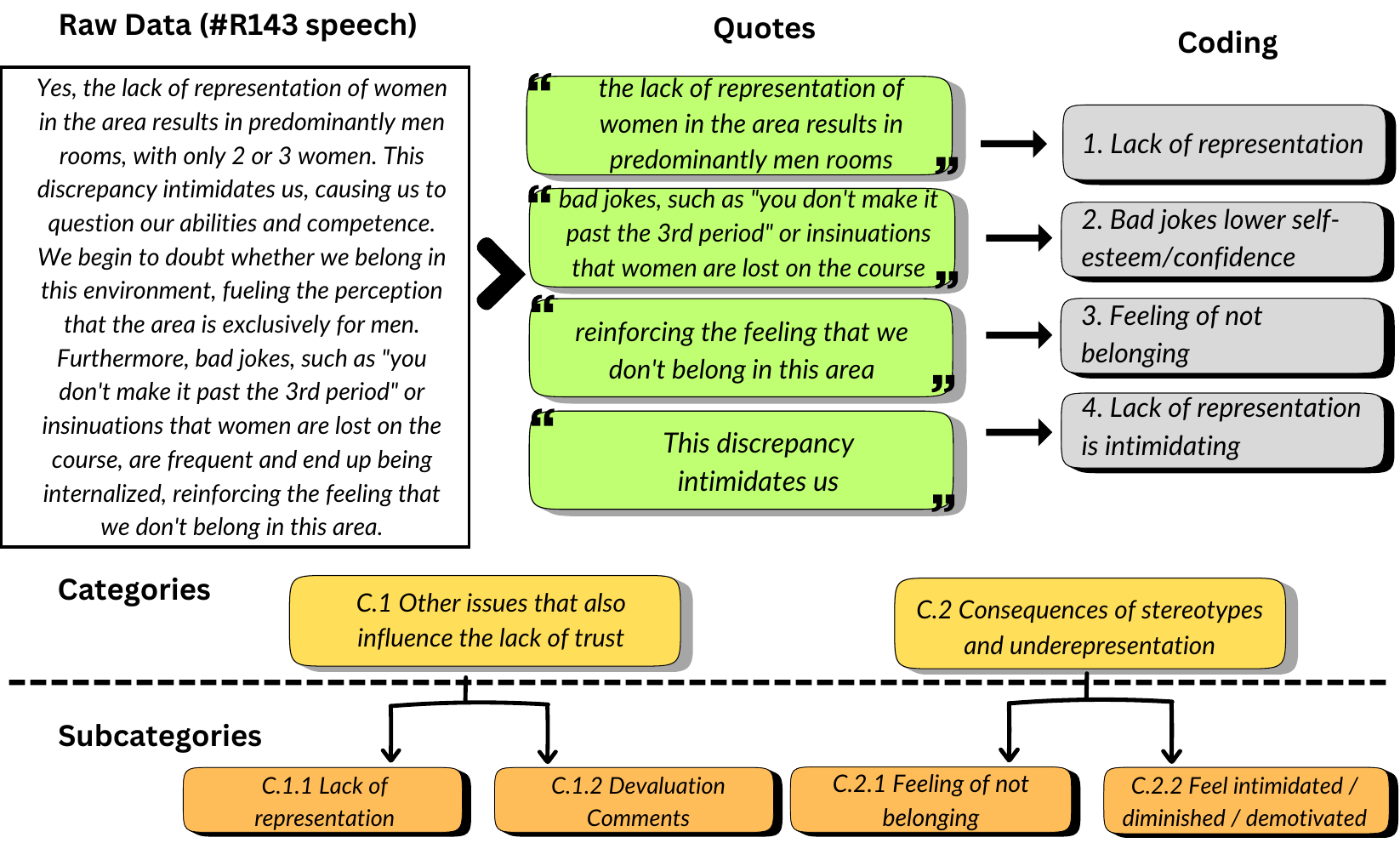} 
     \caption{Coding Process Overview}
     \label{img:coding}
\end{figure}

\textbf{Step 6: Performing the data analysis.} We adopted a mixed-methods approach, combining qualitative research with quantitative analysis through visualizations and correlations.
In the qualitative analysis, we employed procedures of the Grounded Theory by performing open and axial coding~\cite{corbin2008basics} to deepen the understanding of the analyzed data. For the coding analysis, all paper authors contributed to the open coding, in which each one focused on at least one question from the survey instrument. Next, we moved on to axial coding, which was conducted by all authors. This phase involved a more in-depth analysis of the initial categories to establish relationships between the codes and form a more coherent and integrated structure of the identified categories and subcategories. Finally, each open-ended question of Table~\ref{tab:survey} was revised by at least two authors. Figure~\ref{img:coding} exemplifies our coding process. The example shows how the response from participant number \#R143 was analyzed. In this example, this participant commented on how gender stereotypes in the ICT field influence the lack of confidence that some women may experience (Q33). Based on the extracted quotes, the \#R143 response (raw data) was initially subdivided into four codes. The bottom of the figure presents the final categories and subcategories created.
Notice that the same respondent's answer can give rise to different categories and subcategories.

\section{Study Results}\label{NEWresults}

\textbf{Participants characterization.} 422 students responded to the survey. However, 20 of them were not part of the target group and their responses were disregarded. Thus, we considered 402 valid responses. 
We received responses from 18 Brazilian states.
The majority of participants were between 18 and 24 years old, representing 57.2\% of the total, followed by 25 to 34 years old (30.9\%). The other age groups had smaller proportions, with 8.8\% between 35 and 44 years old, 2.7\% between 45 and 54 years old, and only 0.5\% aged 55 to 64 years old (Q4). Additionally, 49.75\% of participants identified themselves as white, 34.08\% as brown, 14.18\% as black, 0.75\% as yellow, and 0.25\% as indigenous; only 1\% preferred not to answer (Q3). Regarding sexual orientation, 65.92\% identified as heterosexual, 20.65\% as bisexual, 6.97\% as homosexual, 2.49\% as pansexual, and 1.24\% indicated other sexual orientations, while 2.74\% preferred not to answer (Q2).

About the survey's undergraduate program data (Q6), Computer Science, Information Systems, and Systems Analysis and Development were the most taken courses by them, each with approximately 19.4\%. They were followed by Computer Engineering (14.9\%) and Software Engineering (12.19\%).
The other courses received fewer than 10 responses 
totaling 14.2\%, including Information Security, Mechatronics Engineering, and Computer Teaching (Q6). Of these, 55.7\% studied at public universities, 44\% at private universities, and only 0.2\% at philanthropic institutions (Q7).

\textbf{Relation between the total number of students and the number of women.} 
Our research uncovered a mutual growth between the overall student count and the proportion of women students. For instance, in classes with 5 to 15 students, half (50\%) of the interviewees pointed out 1 to 3 girls. In turn, for most of the larger classes, the highest representation of women was concentrated in the range of 4 to 6 students. In classes of 40 or more students, a noteworthy 28\% of participants highlighted the presence of 10 to 12 girls, evidencing a discernible rise in the representation of women students (Q9 \& Q10). However, this upward trend is subtle, showcasing a non-linear relationship, since the increment in the number of girls does not directly mirror the growth of the overall student population. 



\subsection{Impact of Classroom Interactions (RQ1)}

More than 80\% of the participants had already heard insinuations that computing-related courses are more appropriate for boys than girls (Q17). This stereotype contributes to a significant barrier for women, manifesting in various forms of gender bias and discrimination within academic environments. 
One of the most pervasive issues faced by women in these settings is the discomfort and intimidation they experience when attempting to express their opinions or engage in discussions. Participant \#R358 said \textit{``I need to spend much more energy than my 
men colleagues to be heard; I have to provide many more arguments for my opinion to be considered"}. This increased effort 
reflects an underlying bias against women's contributions and capabilities. 

The answers shared by women on expressing themselves highlight a pervasive issue of gender-based interruptions in academic settings (Q24). 
For instance, 60.19\% of them reported being interrupted while expressing their opinions or asking questions in the classroom (Q22).
When asked about the individuals responsible for these interruptions (Q23), 51.24\%  indicated 
men peers, 27.36\% pointed to teachers, and 0.74\% 
mentioned other 
men members of the academic community. Only 14.42\% highlighted they had experienced interruptions from women.

These data reveal a common topic where women frequently face interruptions or dismissals from their 
men peers and professors, in contexts where men are the majority. 
In one scenario, 
\#R149 shared the following situation: \textit{``Whenever I answered a question, the professor cut me off and did not look at me, only at my 
men colleague. The first time it happened, I found it strange, but as it continued, [...] I felt increasingly insignificant up there [...]
trying to understand what had happened, whether the problem was personal or related to my gender"}. 
This situation not only silenced her contribution but also implicitly questioned her expertise in the field.


The impact of these interruptions on women's self-esteem and confidence varies (Q19).
Some women find ways to assert themselves despite challenges, learning to speak louder or continue their thoughts after being interrupted. A woman cited \textit{``When I was interrupted, I quickly said <I haven't finished yet> and continued speaking"} (\#R383).
Others, however, report that these experiences have a very negative effect. 
\#R94 cited \textit{``This completely discouraged me from continuing to attend the classes"}. This discomfort is not baseless, but is rooted in repeated experiences of being silenced or overlooked. 
The compounding impact of such incidents often leads women to hesitate to share their insights or opinions during discussions. Specifically, 41.05\% of participants chose not to express their opinions in discussions despite being confident in their correctness (Q21). One woman expressed her perspective by saying, \textit{``I prefer to remain silent most of the time, for not having to witness such situations"} (\#R352). Similarly, 35.6\% of participants revealed discomfort in articulating their opinions during class sessions, while 36.32\% expressed a neutral answer. Conversely, 28.11\% reported feeling comfortable when sharing their opinions in the classroom (Q18). 

\medskip
\noindent
\setlength{\fboxsep}{0.3em}
\fbox{
\begin{minipage}{8.3cm}
        \textbf{RQ1} \textbf{Summary}: 
        Women often face discomfort and intimidation in educational environments, finding it difficult to express their opinions and participate in discussions. Also, they need to put in more effort to be recognized compared to their 
        men peers. It is crucial to promote a culture of respect that values 
        women's contributions equally.
\end{minipage}
}
\medskip

\subsection{Challenges Women Undergraduate Students Face in ICT programs (RQ2)}


Enrollment in ICT undergraduate programs can pose a challenging journey for women. 
We identified 14 categories and 25 subcategories based on the situations mentioned by respondents in questions Q26, Q33, and Q37.
The \textbf{Sexism/Gender Discrimination} category was split into two subcategories, \textit{Discrimination in the Academic environment} and \textit{Sexist society}.
For the first one, participants cited the professors' lack of seriousness and consideration; stereotypes against women in computing; and concerns about conditional favors from professors. For instance, \#R338 and \#R345 respectively mentioned: \textit{``I should be careful because a professor gave me a recommendation letter, and he might want something in return.''}, and \textit{``A professor once asked me how many liters of detergent it takes to wash ten dishes, implying that my perspective was valuable only in domestic contexts.''}. 

About the subcategory \textit{sexist society}, participants highlighted that stereotypes, from childhood, place women in the role of caregivers for the home, husband, and children.
Men are encouraged to think, decide, and act with courage, while women receive less encouragement, leading girls to believe that technology is a male domain. 
This type of prejudice occurs both inside and outside the area and contributes to increasing dropouts from ICT courses, demotivating students and hindering women's advancement.
Along these lines, respondent \#R22 stated 
\textit{``[...] Our ability and intelligence are always put to the test and this is tiring and causes several problems for us''}.

The \textbf{Inferiority Complex and Self-Esteem}
category includes \textit{Feeling of Failure, Incapacity, Insecurity, and Inferiority}. In this context, participants mentioned feelings such as a failure in work environments predominantly made up of men or under the leadership of men; 
and being treated as inferior from their first job. For instance, \textit{``[...] I feel extremely inferior to my colleagues, mainly because I have the impression that I have less previous knowledge [...].''} (\#R262).

The \textbf{Difficulty in the Academic Environment} category has
five subcategories: \textit{General Difficulties}, \textit{Difficulty in Programming Courses}, \textit{Complex or Inaccessible Teaching Approach}, \textit{Difficulty in Hardware Courses}, and \textit{Difficulty in Mathematics Courses}. 
The participants described situations such as failing courses, feeling academically behind peers, struggling with practical programming and hardware tasks, and difficulties understanding specific content. For instance, as mentioned by \#R270 \textit{``I feel like everyone is ahead of me in terms of learning.''},
and \#R98 \textit{``Subjects in the disciplines are presented as if prior knowledge is assumed, challenging novices.''}

The \textbf{Social Expectations and Self-affirmation}, \textbf{Social Perceptions of Gender Inequality}, \textbf{Skill Self-invalidation}, and \textbf{Feeling of Emotional Invalidation} categories 
represent situations related to \textit{Pressure to Prove Capacity and Validate Skills}, \textit{Perceived Underestimation/Need to Validate Women’s Opinions}, \textit{Self-invalidation and Uncertainty about Skills}, and \textit{Impostor Syndrome}. For instance, the \#R11 cited the constant need to prove their abilities to others: \textit{``I feel like I need to be 3 times better than my 
men colleagues.''}. A similar feeling was cited by \#R228: \textit{``They doubted that I was right about something, and only believed me after a man said I was right''}.

According to surveyed women, the \textbf{Underrepresentation of women} makes the academic and professional journey challenging and tiring.
Usually, it leads women to believe that they are unwelcome and that the field is exclusively for men.
For example, \#R153 mentioned \textit{``[...] This discrepancy intimidates us, causing us to question our abilities and competence. We begin to doubt whether we belong in this environment[...]}.
Sometimes, the discrimination comes from the professors (\textbf{Demotivation Promoted by Professors}), through sexist and rude comments and attitudes, even inside the classroom. For instance, in two speeches, the women mentioned that they were (in)directly related to prejudiced statements 
as following: 
\textit{``A friend of mine left a class in tears after the professor repeatedly referred to her, the only woman, while mentioning a washing machine.''} (\#R352) 
and \textit{``I once heard from a professor (man) that the girls in the class should look for a rich man [...] ''} (\#R139).

Another mentioned challenge is the \textbf{Association with another area/subarea}. They had been encouraged to move to areas outside of computing (such as administrative or organizational) or less technical areas of computing, such as, according to them, front-end development and design. Some even stated that men said these other areas of computing are more suitable for women because they are more detail-oriented than men, while the other areas involve more technical skills.
\textbf{Social and Family Issues} was also mentioned by participants.
For instance, \#R220 cited the lack of familiar support: 
\textit{``[...] It's hard to perform well when I manage everything alone—cooking, cleaning, working to support myself—while my colleague enjoys financial stability and a supportive home environment [...]''}.
Along these lines, they commented that domestic activities, which women often carry out, end up reducing the time they have to study compared to men. 

\begin{figure}[ht!]
\begin{tikzpicture}
    \begin{axis}[
        width=5.8cm,
        height = 3.0cm,
        xbar=0.04cm,
        bar width=2.0pt, 
        xmin=0,
        xmax=100,
        font=\tiny,
        nodes near coords,
        ytick=data,
        nodes near coords align={horizontal},
        symbolic y coords={
           Pregnancy during undergraduate studies,
           Racial stereotyping,
           Sexual harassment,
           Discri. behaviors affecting opportunities,
           Lack of mentoring and support,
           Communication difficulties,
           Isolation,
           Bad jokes about being a woman,
           Difficulties in balancing academic and personal life,
           Gender stereotypes regarding technical ability,
        },
    ]
    
    \addplot[fill = pink] coordinates {
        (1.24,Pregnancy during undergraduate studies)
        (8.20,Racial stereotyping)
        (20.14,Sexual harassment)
        (26.11,Discri. behaviors affecting  opportunities)
        (40.04,Lack of mentoring and support)
        (46.76,Communication difficulties)
        (49.75,Isolation)
        (50.49,Bad jokes about being a woman)
        (58.45,Difficulties in balancing academic and personal life)
        (59.70,Gender stereotypes regarding technical ability)
    };      
    \end{axis}
\end{tikzpicture}
\caption{Frequency of Common Challenges}
\label{fig:FrequencyofChallenges}
\end{figure}
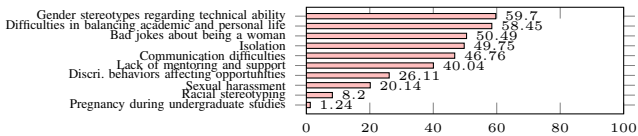

We identified three more categories: \textbf{Lack of Recognition}, \textbf{Abusive and Minimizing Behaviors}, and \textbf{Prejudice}. We describe some situations as follows. The \#R9 stated on the lack of recognition: \textit{``After presenting my undergraduate thesis and achieving a good result, all my colleagues attributed the credit ONLY to my advisor.''}. Additionally, the \#R340 mentioned: 
\textit{``A colleague privately apologized for not acknowledging my help with a problem in a WhatsApp group.''}.
Additionally, 
we also asked the participants whether they had experienced challenges commonly cited in the literature~\cite{doi:10.1080/13668803.2018.1483321, 10.1145/3613372.3613394, 10164701,DBLP:conf/icse/TahsinAAS22}. 
Figure~\ref{fig:FrequencyofChallenges} shows the frequency of specific challenges women may face as undergraduate students (Q32), such as \textit{Gender stereotypes regarding women's technical capacity} (59.70\%), \textit{Difficulties in balancing academic and personal life} (58.45\%), and \textit{Communication difficulties} (46.76\%). 
The percentages together exceed 100\% because each participant could choose more than one challenge.

These results highlight the persistent gender inequality, revealing forms of discrimination and the hostile environment faced by women in ICT programs.
For example, just over half of the participants have heard \textit{Bad jokes} in the academic environment, which may lower self-esteem, make them doubt themselves, and 
ends up contributing to evasion.
Similarly, \textit{Isolation} was reported by nearly half of them (49.75\%), often 
due to the low representation of women in ICT courses. 
This scenario not only worsens the perception of inequality but can also negatively impact students' academic performance and motivation. 
Besides, \textit{Communication difficulties} (46.76\%) may also be associated with this gender dynamic in the ICT field, as 
argued in Section \ref{sec:RQ_interrupcoes}.

Furthermore, some participants disclosed instances of \textit{Sexual harassment} perpetrated by professors or colleagues within the institution (20.14\%). This troubling issue was exemplified by \#R381, who mentioned: \textit{``Some professors have already committed harassment [...] 
and are still teaching subjects I need to take in the future, which makes me very uncomfortable and scared."}. This speech highlights not only the prevalence of harassment, but also the continuous fear and anxiety that it causes among students.
Also, \textit{Racial stereotyping} (8.20\%) emerges as a notable challenge in discussions on inequality. The challenges of race and gender intersect, intensifying experiences of discrimination.
Another challenge highlighted was \textit{Pregnancy during undergraduate studies} (1.24\%).
Balancing motherhood with academic pursuits or employment can pose significant challenges, such as overload, insufficient time for self-care and keeping up to date professionally, difficulties also pointed out by Rocha et al.~\cite{10164701}.


Additionally, the \textit{Lack of mentoring and support} posed a significant obstacle (40.04\%). We also explored initiatives in educational institutions prioritizing diversity and gender equality in ICT (Q35), revealing 38 initiatives (see Table~\ref{tab:initiatives}).
Furthermore, 26\% of participants reported past or current involvement in one of these initiatives. Participation in these programs can be crucial in retaining women in 
fields dominated by men such as ICT.



\begin{table}[ht!]
\centering
\caption{Initiatives to Promote Gender Equality}
\label{tab:initiatives}
\footnotesize
\begin{tabular}{|p{8.5cm}|}
\hline
\rowcolor[HTML]{C0C0C0} \multicolumn{1}{|c|}{\textbf{Initiatives}} \\ \hline
 Array Girls, Besouras Digitais, Boss, Brg Mulher, CIntia, CodeRosa, Django Girls, Elas Programando, Elas@Computacao, Elas++, Enep.gov, Gatech, Girls Power Programming
Grace, Grace Hopper, Grupo Conectadas, IEEE WIE, IT Girls, Lovelaces, Meninas Cientistas, Meninas Digitais, Meninas Digitais do Vale, Meninas dos Sertões de Crateús, Meninas na Ciência, Meninas.comp, MeninasOn, Minas da Linux tips, MulherAda, Mulheres de Anitta, Mulheres na Tecnologia, ProgramAda, Projeto Adas, Projeto CodeUp, Pyaldies, Robchicas, Techgirl Mulheres na TI, Twist, Womakers \\ \hline
\end{tabular}
\end{table}

\medskip
\noindent
\setlength{\fboxsep}{0.4em}
\fbox{
\begin{minipage}{8.2cm}
        \textbf{RQ2} \textbf{Summary}: 
       A woman undergraduate student may face challenges such as lack of recognition for achievements, sexism, feelings of inferiority, communication difficulties, isolation, and lack of support. The findings indicate that these difficulties shape women's confidence and emphasize the need for targeted interventions to promote inclusion and gender equality.
\end{minipage}
}
\medskip

\subsection{Approaches to Improve Women's Confidence in Tech (RQ3)}
\label{subsec:approaches_to_improve_confidence_in_technology}

Of the 402 valid survey answers,
$\sim$93.5\% of participants believe that gender stereotypes contribute to women's lack of confidence (Q33). They highlighted challenges faced by women undergraduate students, creating an environment where they feel intimidated, diminished, demotivated, discouraged, oppressed, and embarrassed, which can undermine their confidence in their abilities and contributions. Table~\ref{confidence} shows the strategies mentioned by them that could be explored to improve women's confidence in technology (Q34).

\begin{table}[ht!]
\centering
\scriptsize
\caption{Strategies to Boost Women's Confidence in Tech}
\label{confidence}
\begin{tabular}{|l|p{5cm}|r|}
\hline
\rowcolor[HTML]{C0C0C0}\textbf{Category}& \textbf{Subcategory}	& \textbf{\# Cited} \\ \hline
Gender Diversity & Leadership positions for women	        & 3 \\ \cline{2-3}
 & Visibility for women    & 10
\\ \cline{2-3}
 & Attracting More Women (Students and Professors)	        & 76 \\ \hline
 Support Program	& Internships specifically for women & 2 \\ \cline{2-3}
& Creating programs for women  & 3 \\ \cline{2-3}
&  Projects for Women  & 4 \\ \cline{2-3}
& Lectures and Outreach Programs & 8 \\ \cline{2-3}
& Mentorship Programs & 14 \\ \cline{2-3}
& Support Groups        & 52 \\ \hline
Social behaviour & Create people’s awareness actions & 31 \\ \cline{2-3}
& Empathy among team members & 3 \\ \hline
Sexist behaviour & Reducing stereotypes and prejudices & 4 \\ \hline
Policies  &  Policies for protection against moral harassment	& 4 \\ \hline
Networking  &  Networking Groups 	&           4 \\ \hline
Code of conduct	& Code of conduct for men        & 1 \\ \hline
\end{tabular}
\end{table}

\textbf{Gender Diversity} was the primary concern for the students, highlighted by \#89. Notably, the subcategory \textit{Attracting more Women} (\#76) received the most mentions.
Participants emphasized the need to include women professors, noting that students feel a greater sense of representation when they have women role models to look up to.  Thus, respondent \#R103 stated \textit{``If more women professors were leading classrooms and projects, it would bring me more confidence [...]
feeling I'm capable too."}
The Gender Diversity category encompasses other subcategories, such as \textit{Visibility for Women (\#10)}. 
According to women students' perspectives, it's crucial to highlight the women who have played significant roles in the history of computing. 
It inspires and encourages other women.
The 2nd most mentioned category by students was the \textbf{Support Program} (\#83). According to them, \textit{Support Groups} could empower women to take on leadership roles and become protagonists. These groups would offer confidence and comfort when classmates, professors, or coworkers challenge their abilities. For instance, \#R266 stated: \textit{``I believe that a strong awareness and support program for women in tech would be essential to strengthen our self-confidence [...]."}

The 3rd most mentioned category was the \textbf{Social Behavior}, with 34 citations. 31 respondents suggested the creation of awareness actions to improve women's confidence in the ICT field.
For instance, participant \#R190 stated: \textit{``It is necessary to raise awareness among professors and classmates about behaviors that exclude, minimize, or harm women's participation in the ICT field. Professors need training regarding avoiding stereotypes [...]
and discrimination-free communication. Professors should not make bad jokes in the classroom, and when someone does, they should intervene}." 

We identified four more categories: \textbf{Sexist behavior, Policies, Networking, and Code of conduct}. These categories were mentioned less frequently by the participants. The respondent \#R190 stated: \textit{``It is necessary to have more projects that promote networking among women who study and work in the ICT field, as well as with partner companies that promote gender equality in their institutions, to promote this equality in the job market as well"}. Surprisingly, only one participant suggested the creation of a code of conduct for men.
This finding differs greatly from the findings obtained by other researchers, such as Trinkenreich et al.~\cite{DBLP:journals/tosem/TrinkenreichWSG22}, who identified that ``Create and enforce a Code of Conduct" is one of the strategies to increase women’s participation in OSS Projects.

\medskip
\noindent
\setlength{\fboxsep}{0.3em}
\fbox{
\begin{minipage}{8.3cm}
  \textbf{RQ3.} \textbf{Summary}: 
Specifically, two main strategies were addressed to promote a more inclusive environment in the ICT area: attracting more women to roles as students and professors, emphasizing the importance of having women role models, 
and raising awareness among professors and peers about discriminatory behaviors.
\end{minipage}
}
\medskip

\section{Discussion}\label{discussion}

The ICT field remains predominantly male-dominated, discouraging many women from pursuing tech careers despite genuine interest. This underrepresentation leads to challenges such as low self-esteem, feelings of inferiority, and a sense of not belonging \cite{DBLP:conf/icse/BomanAN24}. The absence of women role models, particularly professors, reinforces gender stereotypes, further discouraging women from entering or persisting in the field. These dynamics contribute to high dropout rates and a sense of exclusion, limiting women’s engagement and success.

Women in ICT courses face gendered stereotypes, including the need to provide additional arguments to be heard and a higher likelihood of being interrupted by men 
classmates. Such interruptions undermine their contributions and erode confidence, with many women feeling invisible and questioning their abilities. Additionally, reports of discrimination, conditional support from professors, and persistent stereotypes labeling technology as 
men dominated highlight a sexist culture that undervalues women’s skills and contributions.

Participants noted internal challenges, such as feelings of inferiority and self-doubt, that stem from comparisons with 
men classmates and a perceived gap in recognition. Social and family pressures, such as domestic responsibilities and limited support, which hinder their focus on studies. Addressing these issues requires a multi-faceted approach. Early exposure to technology and programming for young girls helps break down stereotypes and fosters interest in ICT careers \cite{doi:10.1080/08993408.2022.2160144,hinckle2020relationship}. Integrating technology into curricula, highlighting successful 
women role models, and establishing supportive learning environments are also critical steps. By encouraging gender-diverse group work, involving 
men professors in supporting 
women students, and initiating institutional projects to address gender issues, educators and institutions can help build a more inclusive environment \cite{DBLP:conf/ecsa/Singh19, dasgupta2014girls}.


The involvement of professors, educational institutions, and public authorities is vital to fostering a supportive academic atmosphere. 
Addressing stereotypes and prejudices within academic settings is key to reducing the gender gap in ICT 
\cite{DBLP:conf/esem/CanedoBOS0M20}. Gendered interruptions and exclusionary behaviours discourage women from staying in tech, perpetuating their underrepresentation. Institutions can foster a culture of respect and recognition by challenging these behaviours, creating an environment where women feel welcomed and supported.
Finally, participants recommended that women actively seek internships, engage in technical communities, prioritize ongoing self-improvement, and build confidence and resilience. This proactive approach enables women to form strong professional identities and thrive in the tech industry.
\section{Threats to Validity}\label{threats}

\textit{External validity}:
We mitigate threats to external validity by analyzing student perceptions across various Brazilian universities, courses, and states. Despite 402 responses from diverse Brazilian states, it's important to note that all respondents are exclusively from Brazil. \textit{Internal validity}:
The characteristics of our sample, consisting entirely of respondents from Brazil, may have influenced our results. Consequently, the challenges and suggestions identified in the study may primarily reflect the experiences and circumstances specific to women students in Brazil. Additionally, it's important to acknowledge that the majority of the authors involved in this research are women, which could also have influenced our findings.
\textit{Construct validity}: 
Before distributing the survey, we conducted three pilots. Even with this review and subsequent care, it is still possible to incur errors, such as questions with similar content that can generate repeated answers. 
 We also recognize that the wording of some questions may indirectly influence the responses (even with the ``never'' option on the Likert scale). 
All researchers participated in the coding process to mitigate the threat of \textit{conclusion validity} related to qualitative analysis. Additionally, each researcher's coding was independently reviewed by another author of the research team.

\section{Conclusions and Future Work}\label{conclusions}

This paper presented the challenges faced by women in ICT programs, marked by a journey filled with isolation, alienation, and discriminatory practices. The pervasive sense of self-doubt can have profound implications for women individuals' careers, academic pursuits, and overall mental health. 
To overcome these challenges, it is vital to boost women's confidence through a comprehensive approach. This includes addressing educational settings, workplace cultures, and societal perceptions. Support and mentoring groups play a key role, in offering solidarity, encouragement, and experience-sharing. 
Future works involve a more in-depth statistical analysis of the data to examine the role of gender stereotypes in shaping perceptions of competence and the lack of confidence often experienced by women in ICT. In addition, we consider to conduct a replication on a global scale. 
Additionally, we plan to investigate the influence of specific situations, such as moments of failure or doubt about skills on women's motivation and academic performance, as well as how early exposure to technology might positively affect their attitudes toward ICT.



\bibliographystyle{IEEEtran}
\bibliography{sample-base}

\end{document}